\documentclass[twocolumn,preprintnumbers,amsmath,amssymb]{revtex4}
\usepackage{graphics}% Include figure files
\usepackage{epsfig}
\usepackage{dcolumn}% Align table columns on decimal point
\usepackage{bm}% bold math
\usepackage{tabularx}

\begin{document}
%\begin{CJK*}{GBK}{song}
%\preprint{APS/123-QED}

\title{The Spin-Orbit term in the Nuclear Shell Model}% Force line breaks with \\

\author{H. M\"{u}ther}\thanks{herbert.muether@uni-tuebingen.de}\affiliation{Institut f\"{u}r Theoretische Physik, Universit\"at T\"ubingen, Auf der Morgenstelle 14, D-72076 T\"ubingen, Germany}

\date{\today}% It is always \today, today,
             %  but any date may be explicitly specified

\begin{abstract}
Quasi-nuclear systems, representing nuclei with variable size, are studied to investigate the occurrence of the spin-orbit term in the nuclear mean field in the transition from infinite nuclear matter to finite nuclei. Relativistic as well as non-relativistic mean field calculations based on models for the nucleon-nucleon ($NN$) interaction, which fit the $NN$ scattering data, are considered. A very strong correlation between the strength of the spin-orbit term and radius of the nuclear system is observed. The origin of the spin-orbit term is analyzed by inspecting the contributions of the different partial waves and various mesons in a One-Boson-Exchange model of the $NN$ interaction. Also results for a realistic interaction model based on chiral effective field theory including the contribution of three nucleon interactions are discussed. The influence of correlation effects and the enhancement of the small component of Dirac spinors for nucleons in the nuclear medium is discussed.
\end{abstract}

%\pacs{24.10.Ht;24.10.Cn;24.10.Jv;21.65.Cd}% PACS, the Physics and Astronomy
                             % Classification Scheme.
%\keywords{Suggested keywords}%Use showkeys class option if keyword
                              %display desired
\maketitle

\section{Introduction}
The nuclear shell model, which describes the nucleus as a system of protons and neutrons moving independently in a mean field generated
by the interaction with all other nucleons, is the starting point of essentially all microscopic nuclear structure studies\cite{ringschuck}.
A rather strong spin-orbit term is an important ingredient of the nuclear mean field. Only after Goeppert-Mayer\cite{Ma49} and Haxel, Jensen and Suess\cite{HJS49} had suggested to incorporate such a spin-orbit term in the phenomenological Hamiltonian for the nuclear mean field, they were able 
to reproduce the empirical magic numbers, which occur in systematic studies of binding energies and nucleon separation energies.

This spin-orbit term is not only required to describe binding energies and separation energies, it is also an important part of the optical model potential for 
nucleon - nucleus scattering\cite{opmo1,opmo2} and is a dominant feature in all microscopic studies of nuclear spectroscopy. As an example we mention
the role of the spin-orbit term in suppressing the proton-neutron pairing in nuclei\cite{bertsch10,pairbox}.

Such a spin-orbit term is a special attribute of many-body systems of finite size and cannot be extracted e.g. from studies of infinite nuclear matter, 
which is frequently considered as a bench mark for microscopic nuclear structure studies. Therefore the spin-orbit term is often added to the nuclear mean 
field or a term is added to the phenomenological nucleon-nucleon ($NN$) interaction which generates a corresponding spin-orbit term in Hartree-Fock
calculations as it is done e.g. in the Skyrme forces\cite{VB72}.

The spin-orbit term is generated in a natural way in relativistic mean field approaches as e.g. in the so-called Walecka model\cite{walecka1,walecka2}.  A characteristic feature of such relativistic models is the self-energy for the nucleons, which contains a large attractive component $U_s$, which transforms like a scalar under Lorentz transformation. In the evaluation of the single-particle energy of the nucleons this attractive contribution is compensated to a large extent by a repulsive component $U_0$, which behaves like the zero-component of a Lorentz vector. If one reduces the corresponding  Dirac equation, describing the single-particle properties of nucleons in a spherical nucleus, to a Schr\"odinger equation, one obtains a spin-orbit term, which can reproduce the empirical data of nuclei (see more detailed discussion below). 

But also non-relativistic studies of nuclei based on realistic models for the $NN$ interaction, i.e. models which describe the experimental data of $NN$ scattering, provide reasonable predictions for the spin-orbit structure of the nuclear mean field. In this case it is the spin structure of the two-body interaction which leads to the spin-orbit term in the single-particle spectrum of nuclei. 

The main aim of the studies presented in this paper is to investigate and compare the predictions of relativistic and non-relativistic approaches for the spin-orbit term in the nuclear mean field of light nuclei. ($A\leq 56$). Special attention will be paid to the dependence of the spin-orbit term on the size of the nucleus by investigating spherical quasi-nuclear systems as a function of their radius. The studies are based on realistic One-Boson-Exchange (OBE) models for the $NN$ interaction developed by Machleidt et al.\cite{Machxxx}. Effects of the different partial waves of the $NN$ interaction as well as the influence of the different mesons considered in the OBE model on the structure of closed and open shell nuclei will be discussed.  For a comparison an interaction model based on chiral effective field theory including terms up to forth order in the chiral expansion has been considered\cite{Machn3lo}. Effects of the corresponding chiral 3-nucleon interaction, expressed in form of a density-dependent effective $NN$ interaction\cite{NorbertK1, NorbertK2}, have been investigated.

After this short introduction section 2 will contain the discussion of approaches which keep track of a relativistic structure nucleon self-energy, while section 3 is devoted to non-relativistic approaches. A comparison of the various approaches is presented in section 4, which also contains the conclusion of the studies.

\section{Relativistic Approach}

At first sight relativistic effects seem to be negligible in nuclear structure calculations. The binding energies of the nucleons are much smaller than the mass of the nucleon and the typical values for the kinetic energy of nucleons bound in nuclei indicate that the velocities of the nucleons are well below the speed of light. The reason for the popularity as well as the success of the relativistic approaches is the feature that the resulting self-energy for the nucleon contains a very attractive term $U_s$,  which transforms like a scalar under a Lorentz transformation, and a term $U_0$ which must be treated like the zero component of a Dirac vector. Inserting these two components into a Dirac equation for a nucleus with spherical symmetry leads to
\begin{eqnarray}
      \left[\vec{\alpha}\cdot\vec{p}+\gamma_0(\textit{M}+U_s(r))+U_0(r) \right]
      \Psi_\nu=\varepsilon_\nu\Psi_\nu~ , \label{eq10}
\end{eqnarray}
where we assume that $U_s$ and $U_0$ are local and depend on the radial coordinate $r$. The Dirac spinors, $\Psi_\nu$ can be written in the form
\begin{eqnarray}
\Psi_\nu(\mathbf r) & = & \left( \begin{array}{c} g_\nu(r) \\ -i f_\nu(r) \sigma\cdot{\bf \hat r}\end{array}\right) {\mathcal Y}_{\kappa_\nu m_\nu} (\Omega ) \nonumber\\
& = & \left( \begin{array}{c} g_\nu(r){\mathcal Y}_{\kappa_\nu m_\nu} (\Omega )  \\ if_\nu(r) {\mathcal Y}_{-\kappa_\nu m_\nu} (\Omega ) \end{array}\right) \label{diracspin1}\,.
\end{eqnarray}
All quantum numbers of the states are expressed in terms of the index $\nu$, which represents a radial quantum number $n_\nu$, the projection quantum number for the total angular momentum $m_\nu$ and the quantum number 
$$
\kappa_\nu = \left(2j_\nu+1)\right)\left(l_\nu-j_\nu\right)\,,
$$
representing the angular momenta. Note, that we are suppressing the isospin quantum numbers. As we are considering light nuclei with equal number of protons and neutrons and ignore the effects of the Coulomb interaction, results are identical for protons and neutrons. The upper and lower spinor components in eq.(\ref{diracspin1}) have different orbital angular momenta $l$ and we introduce the corresponding orbital angular momentum $l_\nu'$ for the same total angular momentum by
$$
l_\nu' = \left\{\begin{array}{cc}l_\nu+1 & \text{for } l_\nu=j_\nu-1/2 \\ l_\nu-1 & \text{for } l\nu=j_\nu+1/2\end{array}\right.
$$ 
and $\kappa_\nu' = - \kappa_\nu$. The spherical harmonics $Y_{lm}(\Omega)$ and the Pauli spinors are coupled to form
$$
\mathcal{Y}_{\kappa_\nu m_\nu} = \sum_{m_l,m_s} C(l_\nu m_l, 1/2 m_s\vert j_\nu m_\nu) Y_{l_\nu m_l}(\Omega) \chi_{1/2 m_s}\,.
$$
The Dirac equation (\ref{eq10}) is solved by expanding the radial functions $g_\nu (r)$ and $f_\nu (r)$ in a discrete basis of spherical
Bessel functions. The wave numbers for this basis are chosen such that this discrete basis is a complete orthonormal
basis in a sphere of radius $D$, which is
chosen to be large enough that the results for the bound single-particle states are independent on $D$. With this
expansion the Dirac equation is rewritten in form of an eigenvalue problem and the eigenvalues ($\varepsilon_\nu = E_\nu + M$) and eigenvectors  are
determined by matrix diagonalization \cite{pollsexpan,fritzc49,fritzth}.

Instead of solving the Dirac equation one can also determine the solutions of positive energy by rewriting the two coupled equations of the Dirac eq. (\ref{eq10}) to form a Schr\"odinger equation 
\begin{eqnarray}
\left[-\frac{\nabla^2}{2{M}}+V_{cent}+V_{ls}(r)\vec{\sigma}\cdot\vec{\textbf{\L}}+V_{Darwin}(r)
\right]\varphi_\nu (\textbf{r})\label{eqschr} \\
      \nonumber =E_\nu\varphi_\nu(\textbf{r}),
\end{eqnarray}
where $V_{cent}$, $V_{ls}$ and $V_{Darwin}$ represent
the Schr\"{o}dinger equivalent central, spin-orbit
and Darwin potentials, respectively.
The potentials in Eq.(\ref{eqschr}) are obtained from the scalar $U_s$ and vector $U_0$ potentials as
\begin{eqnarray}
      \nonumber V_{cent}&=&U_s+\frac{\varepsilon}{M}U_0+\frac{1}{2M}[U_s^{2}-U_0^2], 
%\end{eqnarray}
%\begin{eqnarray}
\\
      V_{ls}&=&-\frac{1}{2{M}rD(r)}\frac{dD(r)}{dr}, \label{eq15}
\\
%\begin{eqnarray}
     \nonumber V_{Darwin}&=&\frac{3}{8{M}D(r)}\left[\frac{dD(r)}{dr}\right]^2-\frac{1}{2MrD(r)}\frac{dD}{dr} \\
     &&\nonumber -\frac{1}{4{M}D(r)}\frac{d^2D(r)}{d^2r},
\end{eqnarray}
where  $D$
is defined as
\begin{eqnarray}
       D(r)=M+\varepsilon+U_s(r)-U_0(r)\,. \label{eq17}
\end{eqnarray}
The radial wavefunctions $\varphi_\nu(r)$ resulting from the Schr\"odinger equation (\ref{eqschr}) are related to the corresponding upper component of the Dirac spinors $\Psi_\nu (r)$ in (\ref{diracspin1}) by
$$
\varphi_\nu(r) \sim \frac{g_\nu(r)}{D(r)}\,.
$$
One of the aims of the present study is to investigate the occurrence of the spin-orbit term in the transition from nuclear matter to finite nuclei. For that purpose a set of "quasi-nuclear systems" has been constructed to exhibit this transition e.g. for quasi-nuclear $^{16}O$\cite{pairing}. In this case a sequence of Woods Saxon potentials
\begin{equation}
V_{WS}(r) = \frac{V_0}{1+e^{(r-r_0)/a}}\,.\label{woodsax}
\end{equation}
Assuming a value of $a=0.5$ fm for the surface width, the parameter for the depth of the potential $V_0$ has been adjusted in such a way that for different values of $r_0$ the energy of the first excited single-particle state with $l=0$ occurred at zero energy. Occupying the corresponding $0s$ and $0p$ states with protons and neutrons one obtains a nuclear density distribution with root mean square radii $<r> (r_0)$. These density distributions
\begin{equation}
\rho_{<r>} (r)\,,\label{rhoofr}
\end{equation}
as well as the corresponding single-particle wavefunctions have been used to explore the occurrence of the spin-orbit term due to the localization of the quasi-nuclear system $^{16}O$. 

In an analogous way quasi-nuclear systems for $^{40}Ca$ have been constructed. In this case the depth parameter $V_0$ in eq.(\ref{woodsax}) has been adjusted to localize the $1p$ single-particle state at zero energy and $0s$, $0p$, $0d$ and $1s$ states have been occupied to obtain nuclear density distributions for $^{40}Ca$ with various radii.

As a first attempt the spin-orbit term has been evaluated using the Improved Local Density Approximation (ILDA), which has recently been proposed by Sun {\it et al.}\cite{ILDA}. The ILDA is based on Dirac Brueckner Hartree Fock (DBHF) calculations\cite{subtrt}, which determine the relativistic components of the nucleon self-energy in nuclear matter
\begin{eqnarray}
U_s^{\rm NM}( \rho, \beta, E ) \quad\mbox{and}\quad U_0^{\rm NM} ( \rho, \beta, E) \label{eq:ILDA1}
\end{eqnarray}
depending on the density $\rho$, the proton-neutron asymmetry $\beta$ and the nucleon energy $E$ relative to the corresponding Fermi energy. Since  only isospin symmetric systems will be considered here, the differences of the self-energy terms for protons and neutrons has been dropped and the limit of symmetric nuclear matter ($\beta = 0$) will be considered. The DBHF calculations are based on the Bonn potential\cite{Machxxx} and use the subtracted T-matrix approach\cite{subtrt} to extract the relativistic components. A convenient parameterization of the self-energies of eq.(\ref{eq:ILDA1}) has been presented in \cite{ILDA}.
 
Using e.g. the density profiles defined in eq.(\ref{rhoofr}) one can evaluate local self-energy components 
\begin{equation}
U_{s(0)}^{\rm LDA} (r, E) = U_{s(0)}^{\rm NM} (\rho(r), E)\,.\label{eq:ILDA2}
\end{equation}
The studies of Sun {\it et al.}\cite{ILDA} showed, that this simple Local Density Approximation misses an important surface effect, which is due to the finite range of the interaction. Therefore they used an Improved Local Density Approximation (ILDA)\cite{Jeu77,jamin82,ruirui2}
\begin{eqnarray}
U_{s(0)}^{\rm ILDA}({r, E}) & = & \frac{1}{(t\sqrt{\pi})}\int U^{\rm LDA}_{s(0)}({r'}, E)\Bigl\{\exp\left[-\frac{( r-r')^2}{t^2}\right]\nonumber\\
&& - \exp\left[-\frac{( r+r')^2}{t^2}\right]\Bigr\}\frac{r'}{r}\, dr'\,,\label{eqt1}
\end{eqnarray}
where $t$ is an effective range parameter, which has been fitted to take values
\begin{equation}\label{eqpret}
t=1.3528-0.1322A^{1/3} \quad{\rm [fm]}\,.
\end{equation}
depending on the mass number $A$ of the nucleus under consideration. 

Using the density distributions of eq. (\ref{rhoofr}) the corresponding $U_s^{\rm ILDA}$ and $U_0^{\rm ILDA}$ can easily be calculated and inserted into the Dirac equation (\ref{eq10}) to obtain the resulting single-particle energies $E_\nu = \varepsilon_\nu - M$. Note that this must be done in an iterative way to obtain self-consistent solutions, for which the energy variable $E$ in the self-energies $U_{s(0)}^{\rm ILDA}$ corresponds to the solutions $E_\nu$ of the Dirac equation. This energy-dependence reflects the treatment of the $NN$ correlations in the DBHF calculations, the basis of the ILDA approach. 

\begin{figure*}[htbp] \centerline{\includegraphics[width = 4.5in]{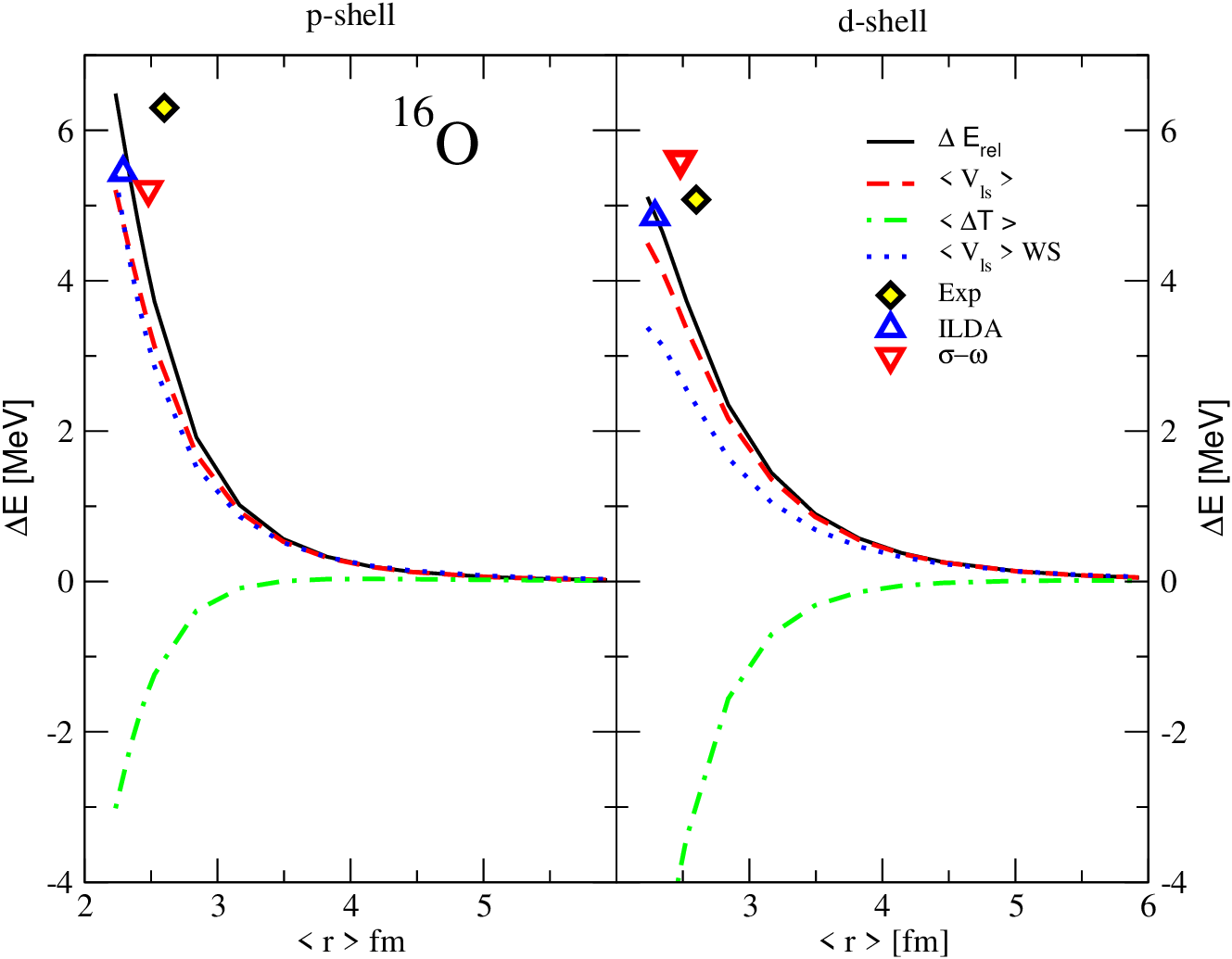}}
\caption{(color online)
 Results for the spin-orbit splitting in quasi-nuclear $^{16}O$ are displayed as a function of the radius of the nucleon density distribution $<r>$. The left panel presents results for the splitting in the $p$-shell (see eq.(\protect{\ref{deltals}})) whereas the right panel shows corresponding results for the $d$-shell. Further explanations in the text.}
\label{fig1}
\end{figure*}

Results for the spin-orbit splitting in quasi-nuclear systems with proton number $Z=8$ and neutron number $N=8$, i.e. quasi-nuclear $^{16}O$, are presented  in Fig.~\ref{fig1} as a function of the radius $<r>$ of the nucleon density distribution $\rho_{<r>} (r)$ of eq. (\ref{rhoofr}). The differences in the single-particle energies
\begin{equation}
\Delta E_{rel} = E_{0p1/2} - E_{0p3/2}\,,\label{deltals}
\end{equation}
obtained from the solution of the Dirac equation are represented by the black solid curve in the left panel of this figure, while corresponding results for the difference between the energies of the $d_{3/2}$ and $d_{5/2}$ shells are given in the right panel of Fig.~\ref{fig1}. 

The results displayed in this figure show a very strong dependence of the spin-orbit splitting on the radius of the underlying nucleon distribution. The spin-orbit splitting disappears for larger radii, which means it occurs only for sufficient localization of the nuclear structure. This effect is larger for the $p$-shell than for the $d$-shell, which represents states above the Fermi energy of $^{16}O$.

In a complete self-consistent ILDA calculation\cite{ILDA} the nucleon density profile to determine the Dirac self-energies are determined from the resulting Dirac spinors. It is worth noting that the spin-orbit terms and radius of such a self-consistent calculation of $^{16}O$, indicated by the blue triangles, denoted by "Dirac" in Fig.~\ref{fig1} are in line with the ILDA calculations using the various nuclear distributions derived from Woods Saxon potentials. This supports the idea that the family of density distributions discussed above is a reasonable choice to explore the dependence of the spin-orbit term on the size of the nuclear system. One may also conclude, however, that the values of the spin-orbit splitting are not very sensitive to details of the density profile.

The experimental data for radius and spin-orbit splitting are represented by the black diamonds in Fig.~\ref{fig1}. The calculations are in reasonable agreement with the experimental data. The calculated values for the spin-orbit splitting and/or the radius of nucleon distribution are slightly smaller than the experimental data. It is a well known feature of DBHF calculations for finite nuclei that they tend to yield density profiles with too small radii\cite{dalrev}. This feature may be a bit enhanced in the present study as the effects of the Coulomb repulsion between protons have been ignored.

In order to study the origin of the spin-orbit splitting in the framework of relativistic mean field calculations, the transformation of the Dirac equation to the Schr\"odinger equation (\ref{eqschr}) has been considered and the expectation values of the spin-orbit potential (\ref{eq15}) 
\begin{equation}
\langle \varphi_\nu \vert V_{ls} \vert \varphi_\nu \rangle\,,\label{eq:expectls}
\end{equation}
have been calculated. The results obtained from these expectation values are plotted as red dashed lines, identified as $<V_{ls}>$ in Fig.~\ref{fig1}. These results derived from the expectation values are very close to the results extracted from the Dirac single-particle energies ($\Delta E_{rel}$ in this figure. Therefore one may  conclude that it is simply this spin-orbit term, which determines the final spin-orbit splitting with high accuracy. 

In detail, however, the situation is a bit more involved: If one determines the eigenstates of the single-particle Hamiltonian using the ILDA for the different nuclear distributions one obtains different radial wave-functions for the states with $j=l+1/2$ and $j=l-1/2$. The $j=l+1/2$ state has a larger binding energy than the one with $j=l-1/2$ and therefore is more localized, which leads to larger kinetic energy. This means that the difference of the corresponding expectation values for the kinetic energies
\begin{equation}
<\Delta T > = <T>_{j=l-1/2} - <T>_{j=l+1/2}\,,
\label{eq:deltat}
\end{equation} 
is negative. The corresponding results are represented by green dashed-dotted lines with label $<\Delta T>$ in Fig.~\ref{fig1}. It turns out that the absolute values of these differences are almost of the same size as the spin-orbit splitting due to eq.(\ref{eq:expectls}). It is interesting to note that this effect in the kinetic energy is compensated with high accuracy by the difference in the expectation values for the central term and the Darwin term of the Hamiltonian, which leads to 
\begin{equation}
\Delta E_{rel} \approx <V_{ls}>\,,
\end{equation}
observed in Fig.~\ref{fig1}.

The blue dotted lines in this figure, labeled as $<V_{ls}>$WS in the legend of the figure, represent the expectation value of the spin-orbit potential (\ref{eq15}) using the wave functions of the Wood Saxon potential (\ref{woodsax}) which has been used to generate the corresponding density distribution with radius $<r>$. Note that the potential (\ref{woodsax}) is a pure central potential, which implies that the radial wave functions for the two spin-orbit states are identical. The results for  $<V_{ls}>$WS are close to the
corresponding expectation values $<V_{ls}>$ which were calculated using the eigenfunction of the single-particle Hamiltonian. This indicates that expectation value $<V_{ls}>$ is not very sensitive to details of the wave functions.

\begin{figure}[htbp] \centerline{\includegraphics[width = 2.8in]{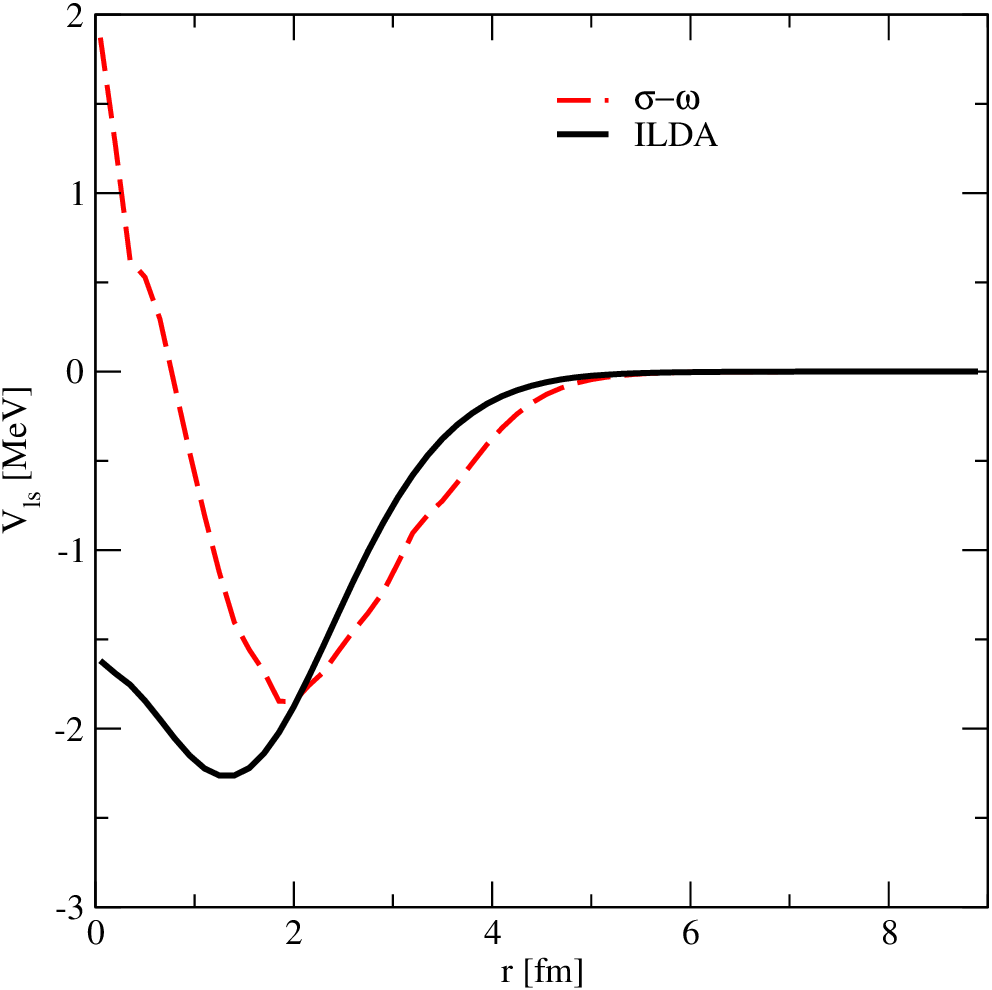}}
\caption{(color online) Radial shape of the spin-orbit potential $V_{ls}$ of eq.(\ref{eq15}). Results for $^{16}$O are presented using the self-consistent ILDA approach (solid line) and the $\sigma-\omega$ model with density-dependent coupling constants derived from DBHF calculations\cite{fritzc49}. }
\label{fig2}
\end{figure}

Fig.~\ref{fig1} also shows results from two self-consistent calculations of $^{16}$O, which are based on DBHF results of nuclear matter using two different Local Density Approximations. The ILDA approach\cite{ILDA} has been introduced above and the results for radius and spin-orbit splitting represented by a triangle with upward orientation have been mentioned above. The second approach has been defined in \cite{fritzc49} extracting density-dependent coupling constants of a mean-field model with a scalar ($\sigma$) and vector meson ($\omega$) to reproduce the Dirac components of the nuclear self-energy derived from DBHF calculations\cite{brockm} of nuclear matter. The radius resulting from this $\sigma-\omega$ model (represented by a downward oriented triangle) is slightly larger than the ILDA prediction but also too small compared with experiment. 

The spin-orbit potentials of eq.(\ref{eq15}) derived from these two self-consistent Dirac calculations are displayed in Fig.~\ref{fig2}. One finds substantial differences for small $r$, which are not relevant in calculating expectation values for orbits with $l=1$ and $l=2$. The differences at the surface of the nucleus reflect the different radii and the fact that the $\sigma-\omega$ model yields a value for the spin-orbit splitting in the $d$-shell, which is slightly larger as compared to the result evaluated within ILDA.

\begin{figure*}[htbp] \centerline{\includegraphics[width = 4.5in]{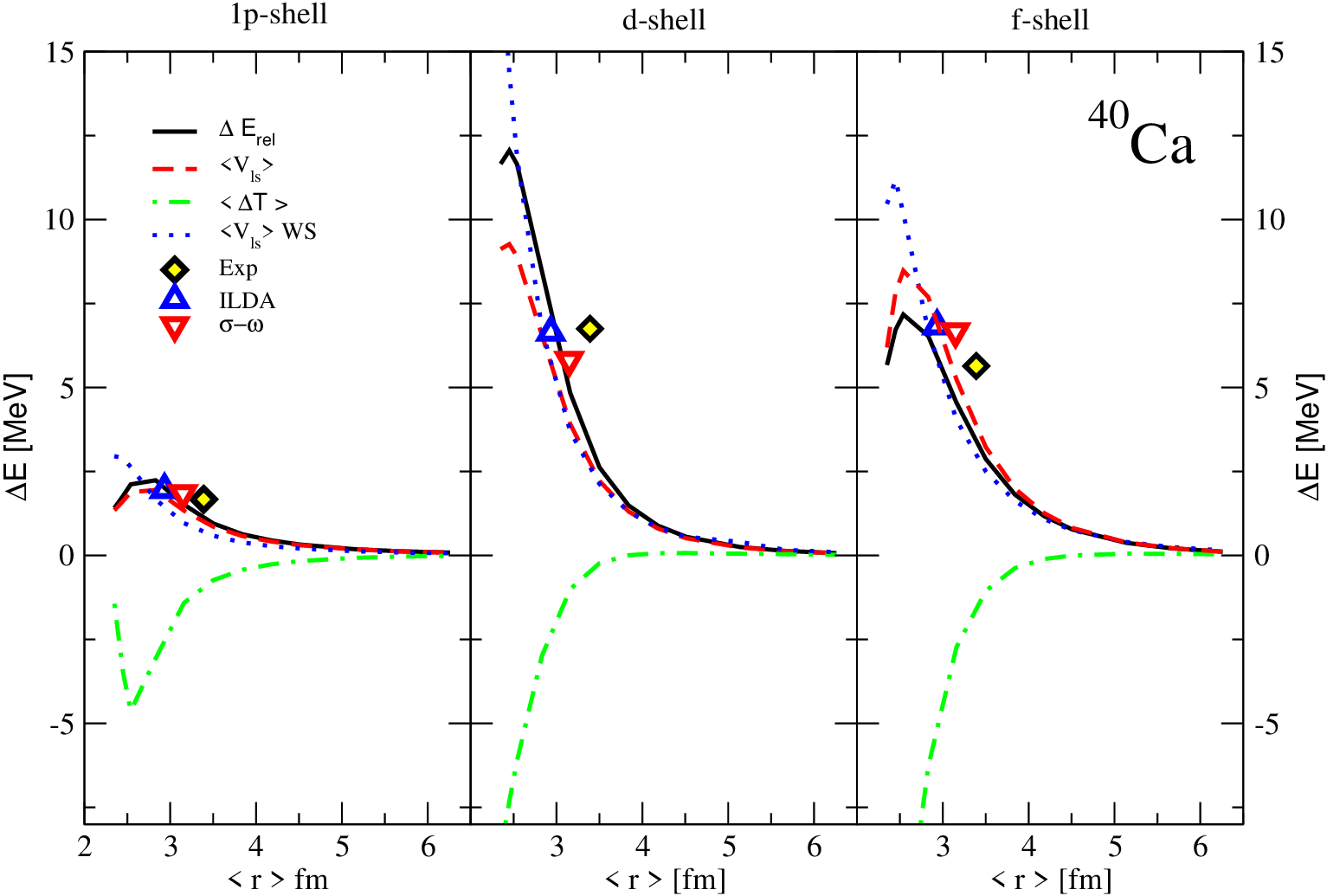}}\
\caption{(color online) Results for the spin-orbit splitting in quasi-nuclear $^{40}$Ca are displayed as a function of the radius of the nucleon density distribution $<r>$. The left, the middle and the right panel present results for the splitting in the $1p$-shell, the $0d$-shell and the $0f$-shell, respectively. The various approximations are discussed in the text. (see also Fig.~\ref{fig1}) }
\label{fig3}
\end{figure*}

Results for the nucleus $^{40}$Ca are displayed in Fig.~\ref{fig3} for the hole states of the $0d$-shell and the particle states of $1p$ and $0f$-shells. The conclusion drawn from the discussion of the results for $^{16}$O, displayed in Fig.~\ref{fig1}, are confirmed by inspecting the results for $^{40}$Ca. The experimental data for spin-orbit splitting of the hole states is larger than the spin-orbit splitting for the particle states. This is non-trivial keeping in mind that $l\dot s$ is larger for the $f$-shell than for the $d$-shell. This trend of the experimental data is nicely reproduced by the self-consistent ILDA and $\sigma-\omega$ calculations and is also visible in study of density distributions of different radii.

The results for spin-orbit splittings in $^{16}$O and $^{40}$Ca derived from the self-consistent ILDA calculations as well as the $\sigma-\omega$ model using density dependent coupling constants are also displayed in table~\ref{tab1}. These results are complemented by results for nuclei with $N=Z$ and closed sub-shells: $^{12}$C, $^{28}$Si, and $^{56}$Ni. Also these calculations have been performed assuming spherical symmetry. It should be noticed that the spin-orbit splitting derived from such calculations tend to predict larger values than derived from experimental data.

\begin{table}
\caption{\label{tab1} Spin-orbit splitting for various nuclei with closed shells or subshells calculated in various approximations (see first column and discussion in  the text) assuming spherical symmetry. The shells under consideration are indicated in the second line. Experimental data for nuclei with closed subshells (numbers in brackets) have been derived from the spectrum of nuclei with one additional neutron. All entries are given in MeV.} 
\begin{ruledtabular}
\begin{tabular}{c|ccccccc}
 &$^{12}$C &\multicolumn{2}{c}{$^{16}$O} & $^{28}$Si & \multicolumn{2}{c}{$^{40}$Ca }& $^{56}$Ni \\ 
 & 0p & 0p & 0d &  0d & 0f & 1p & 0f \\
\hline
 &&&&&&&\\
 Exp & (2.83) & 6.30 & 5.08 & (0.80) & 5.64 & 1.67 & (2.23)\\
 &&&&&&&\\
\hline
 &&&&&&&\\
 ILDA & 4.18 & 5.44 & 4.85 & 5.85 & 6.83 & 1.96 & 6.79 \\
$\sigma-\omega$ & 6.46 & 5.21 & 5.60 & 7.19 & 6.66 & 1.85 & 7.45 \\
 &&&&&&&\\
\hline
 &&&&&&&\\
HF +3N & -2.34 & 3.91 & 4.57 & -0.69 & 4.96 & 1.30 & 0.11 \\
m* & 0.25 & 5.48 & 6.11 & 1.74 & 6.66 & 1.59 & 2.30 \\
&&&&&&&\\
\end{tabular}
\end{ruledtabular}
\end{table}

\section{NONRELATIVISTIC STUDIES}
The aim of this section is to discuss the spin-orbit term of the nuclear shell model within the framework of non-relativistic many-body calculations based on realistic $NN$ interactions. In this case the origin of the spin-orbit term should be related to the spin-structure of the $NN$ interaction, reflected in the spin-dependence of the $NN$ scattering phase shifts. Traditional models of realistic $NN$ interactions like the local interactions of the Argonne group\cite{argv14} or the various One-Boson-Exchange potentials (OBEP) of the Bonn (Idaho) group\cite{Machxxx, erkelenz, holinde} contain strong short range and tensor components, which make it inevitable to employ non-perturbative approximation schemes for the solution of the many-nucleon system. 

One way to get rid of the short-range or high-momentum components of such interactions is to use renormalization techniques\cite{bog1,bog2,bozek,bog3,erik02} to separate 
low momentum and high momentum components of the $NN$ interaction. For that purpose we consider the two-nucleon problem using the Bonn A interaction defined in \cite{Machxxx} and define projection operators $P$ and $Q$ projecting on the subspace of two-nucleon states with momenta below a cutoff $\Lambda$ and the complement, respectively. Using the Unitary-Model-Operator approach (UMOA)\cite{Suzuk} one can define a unitary transformation $U$ in such a way that the transformed Hamiltonian does not couple the $P$  and $Q$ subspaces, which means 
\begin{equation}
Q\,U^{-1}\,H\,U\,P = 0\,,
\end{equation} 
with the original Hamiltonian $H=T+V$ containing the term for the kinetic energy $T$ and the OBE potential $V$. This leads to an effective Hamiltonian
\begin{equation}
H_{eff} = T + V_{lowk}\,,\label{eq:heff}
\end{equation}
with 
\begin{equation}
V_{lowk} = U^{-1} ( T + V ) U - T\,.\label{eq:vlowk}
\end{equation}
The eigenvalues, which are obtained by diagonalizing the effective Hamiltonian of (\ref{eq:heff}) in the P-space, are identical to those, which are obtained in the diagonalization of the original Hamiltonian $H = T + V$ in the complete space. This implies that $V_{lowk}$ yields the same $NN$ phase shifts for nucleons with momenta below the cut-off $\Lambda$ than the original OBE interaction $V$.

If the cutoff $\Lambda$ is appropriately chosen, i.e. around $\Lambda$ = 2 fm$^{-1}$, the resulting low momentum interaction
 $V_{lowk}$ will describe the experimental data up to the pion threshold. Moreover, a very attractive aspect is that this $V_{lowk}$ interaction turns out to be independent of the underlying realistic interaction $V$\cite{bozek}. Uncertainties due to the different models of the high-momentum components of traditional realistic $NN$ interaction have been removed by the renormalization procedure leading to $V_{lowk}$.

With respect to the present study it is a major advantage that $V_{lowk}$ yields rather stable results in lowest order many-body calculations. As will also be demonstrated below the results obtained in mean-field approximation are not very much modified including effects of correlations. 

A major drawback in using $V_{lowk}$  is the fact that it does not provide realistic saturation properties. Using $V_{lowk}$ e.g. in a calculation of nuclear matter, one obtains a binding energy per nucleon increasing with density in a monotonic way\cite{kuckei}. Therefore 3-nucleon forces have to be added to provide good results for the saturation of nuclear matter\cite{bog2} and bulk properties of finite nuclei\cite{goegel}.

This is in line with the use of interaction models based on chiral perturbation theory\cite{weise,machchi}. Also these chiral interactions are limited to nucleons with low momenta and 3-nucleon forces are required to reproduce the saturation properties of nuclear systems (see e.g. the review \cite{samaru} and references there).  This point will further be discussed below.

The first approach to be discussed in this section is based on the Wood-Saxon wavefunctions generated to describe quasi-nuclear systems of varying size representing $^{16}$O and $^{40}$Ca (see eq.(\ref{woodsax})). Denoting the eigenstates of the Wood-Saxon potential by $\vert \mu \rangle$ and $\vert \nu \rangle$ corresponding mean fields and single-particle energies 
\begin{equation}
\varepsilon_{\mu}^{WS} = \langle \mu \vert T \vert \mu \rangle +\sum_{\nu < F} \langle\mu\nu\vert V_{lowk}\vert \mu\nu \rangle \,,\label{eq:epswoods}
\end{equation}
can be calculated with the restriction of the summation on the right-hand side of this equation to states $\nu$ below the Fermi surface of the nucleus considered. 

It is worth noting that the Woods Saxon wavefunctions are expanded in in the same discrete basis of spherical Bessel functions, which has also been used for the solution of the Dirac equation discussed in the previous section. Since the effective interaction is evaluated in a basis of momentum eigenstates in partial waves of the relative basis one has to transform the matrix elements to the plane wave states in the laboratory system using the vector brackets as described in \cite{pw1,pw2}.  This transformation is a bit more involved than the corresponding Talmi-Moshinsky transformation\cite{mosh1,mosh2} to be used for a basis system of oscillator eigenstates.

\begin{figure*}[htbp] \centerline{\includegraphics[width = 4.5in]{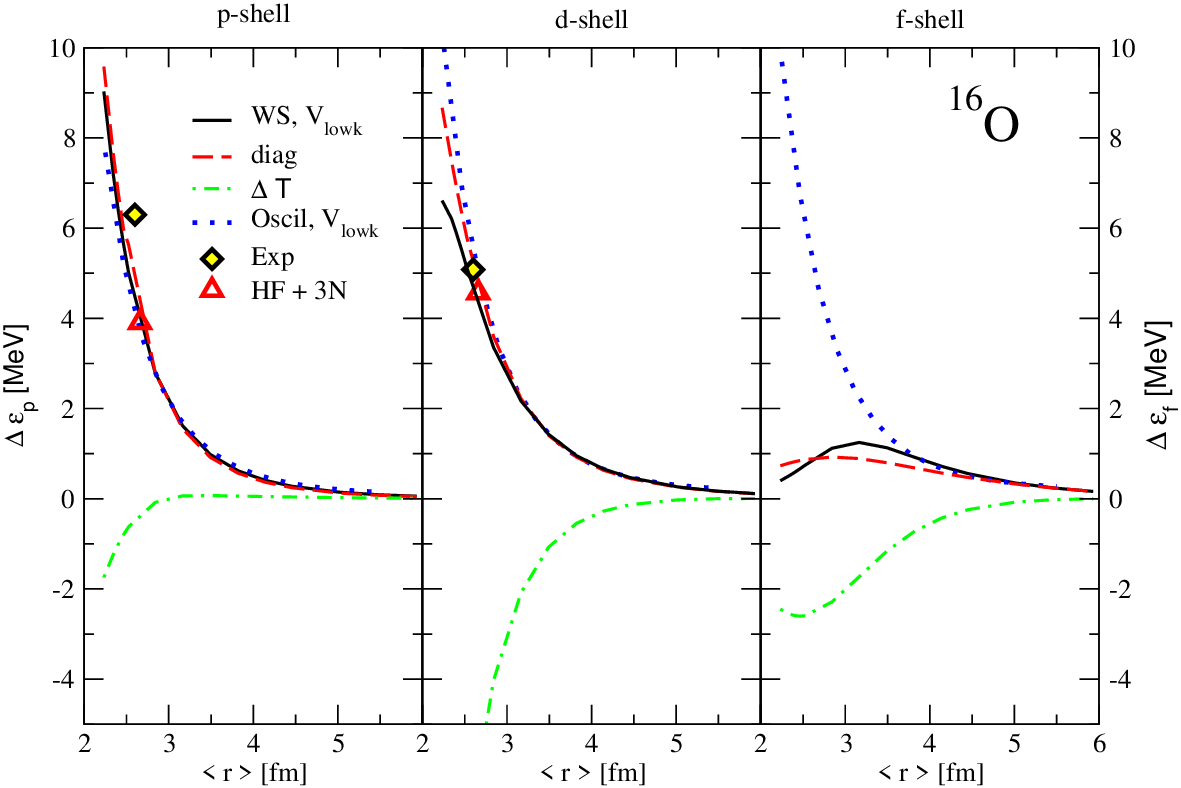}}
\caption{(color online) Results for the spin-orbit splitting in quasi-nuclear $^{16}O$ are displayed as a function of the radius of the nucleon density distribution $<r>$. The left, the middle and the right panel present results for the splitting in the $0p$-shell, the $0d$-shell and the $0f$-shell, respectively. The various approximations are discussed in the text }
\label{fig4}
\end{figure*}

Results for the spin-orbit splitting in the $0p$, $0d$ and $0f$ shell calculated from the energies of (\ref{eq:epswoods}) are displayed in the various panels of Fig.~\ref{fig4} as a function the radius of the quasi-nuclear system $^ {16}$O. The $ls$ splittings using the wavefunctions of the corresponding Woods Saxon potentials are represented by the solid black line using the label ``WS $V_{lowk}$''. The results for the spin-orbit splitting depend very strongly on the radius of the quasi-nuclear system. This is very similar to the results obtained within the relativistic mean field calculations discussed in the preceeding section (see Fig.~\ref{fig1}). For a given radius of the nucleon distribution, however, the results derived from relativistic mean field calculations are typically around 30 percent smaller than the corresponding results evaluated from the $NN$ interaction.

The results are rather insensitive to the details of the single-particle wave functions leading to the same radius for the nuclear system. If one replaces e.g. the Woods Saxon wavefunctions in eq. (\ref{eq:epswoods}) by corresponding eigenfunctions of the harmonic oscillator with varying oscillator length, one obtains the results which are represented by the blue dotted lines (label ``Oscil $V_{lowk}$''), which, in the case of the $0p$ and $0d$ shell, are rather close to the results using Wood Saxon wavefunctions. Differences occur in the case of the $0f$ shell. In this case the Wood Saxon potential yields continuum states whereas the oscillator model generates bound states, which are not very realistic.

In order to test the sensitivity of the spin-orbit energy differences on details of the single-particle wavefunctions one may also consider the single-particle Hamiltonian
\begin{equation}
\langle\kappa\vert h \vert\nu\rangle = \langle \kappa \vert T \vert \mu \rangle +\sum_{\nu < F} \langle\kappa\nu\vert V_{lowk}\vert \mu\nu \rangle \,,\label{eq:epsdiag}
\end{equation}
and derive the single-particle splitting from the eigenvalues of $\langle\kappa\vert h \vert\nu\rangle$. The results are shown in Fig.~\ref{fig4} terms of the red lines with long dashes 
(label ``diag''). In contrast to wavefunctions derived from Wood Saxon or oscillator model, these eigenstates yield different expectation values for the kinetic energies of the states of a spin-orbit doublet. In fact, the differences are non-negligible and provide negative contributions to the energy differences (see green, dashed-dotted lines in Fig.~\ref{fig4}, label $\Delta T$). This difference is counterbalanced by more attractive contributions of the central field in the case of $j=l+1/2$ as compared to $j=l-1/2$. Therefore the resulting spin-orbit splittings almost coincide with corresponding results using Wood Saxon or oscillator states. This is very similar to the phenomenon displayed in Fig.\ref{fig1} discussed above.

\begin{figure*}[htbp] \centerline{\includegraphics[width = 4.5in]{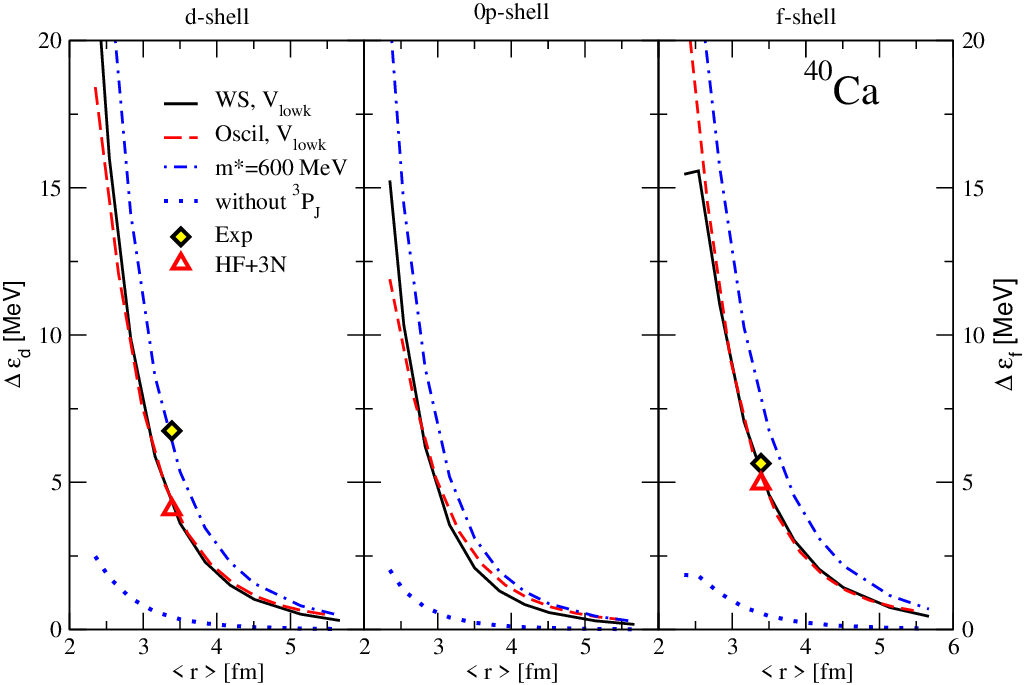}}\
\caption{(color online) Results for the spin-orbit splitting in quasi-nuclear $^{40}$Ca are displayed as a function of the radius of the nucleon density distribution $<r>$. The left, the middle and the right panel present results for the splitting in the $0p$-shell, the $0d$-shell and the $0f$-shell, respectively.  }
\label{fig5}
\end{figure*}

Fig.~\ref{fig4} also displays results for spin-orbit splitting and nuclear radius obtained in a self-consistent Hartree-Fock (HF) calculation using $V_{lowk}$. As it has already been mentioned above such Hartree-Fock calculations do not show saturation in infinite nuclear matter and predict finite nuclei with very small radii. Therefore the $V_{lowk}$ interaction has been supplemented in \cite{goegel} by a 3N force of zero range. leading to a results for the saturation point of nuclear matter as well as  radii and binding energies of light nuclei, which are in good agreement with the experimental data. Results for radius and spin-orbit splitting in $^{16}$O from HF calculations using $V_{lowk}$ supplemented by such a 3N force are represented by red triangles in Fig.~\ref{fig4}. The results are rather close to the corresponding values using Woods Saxon or oscillator functions leading to the same radius for the mass distribution.

The main conclusions resulting from this discussion of results for $^{16}$O displayed in Fig.~\ref{fig4} are confirmed by corresponding results for $^{40}$Ca presented in Fig.~\ref{fig5}.

Results of HF calculations for spin-orbit splittings for various nuclei with closed shells or subshells are also listed in table~\ref{tab1} in the line with the label ``HF+3N''. Comparing the results with those obtained in self-consistent relativistic mean field calculations also listed in this table (ILDA and $\sigma-\omega$ see discussion above) one finds that the results for $^{16}$O and $^{40}$Ca, the nuclei with closed major shells, are rather similar. For nuclei with closed subshells, $^{12}$C, $^{28}$Si, and $^{56}$Ni, however, the situation is quite different. While the relativistic mean field calculations for  $^{12}$C and $^{28}$Si yield positive values for the spin-orbit splitting of the states close to the Fermi energy, the corresponding results of the non-relativistic HF+3N approach yields negative values for the spin-orbit splitting of the $p$ and $d$ shells, respectively. This implies that the assumption of spherical symmetry does not lead to a consistent solutions since e.g. in the case of $^{28}$Si, the energy of the unoccupied $d_{3/2}$ shell is below the energy of $d_{5/2}$, which is assumed to be occupied. This implies that deformed solutions or configuration mixing has to be considered to obtain consistent solutions for these nuclei using realistic $NN$ interactions.

\begin{figure*}[htbp] \centerline{\includegraphics[width = 4.5in]{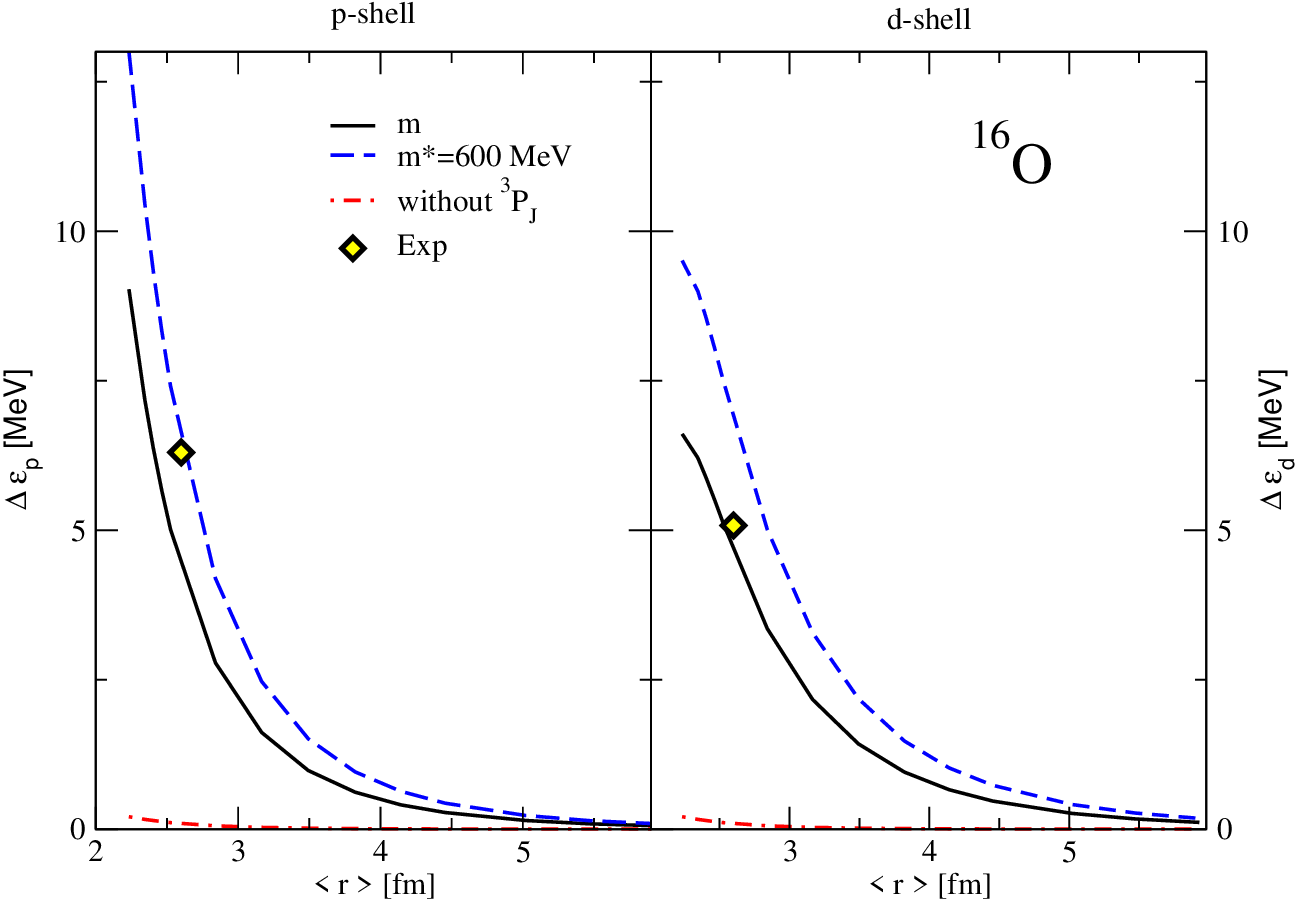}}\
\caption{(color online) Results for the spin-orbit splitting in quasi-nuclear $^{16}O$ are displayed as a function of the radius of the nucleon density distribution $<r>$. The left and  the right panel present results for the splitting in the $0p$-shell and the $0d$-shell, respectively. Results obtained for the full $V_{lowk}$ are compared to those, where contributions from $^3P_J$ partial waves are ignored. The lines with label $m^*$ = 600 MeV are obtained evaluating the underlying OBE potential with Dirac spinors for the nucleon with enhanced small components  }
\label{fig6}
\end{figure*}

The remaining part of this section is devoted to the discussion of the sources  within realistic $NN$ interactions causing the spin-orbit splitting in the mean field of nuclei. For that purpose Fig.~\ref{fig6} presents results for the quasi-nuclear systems of $^{16}$O using modifications of the underlying $NN$ interaction. As a reference the black solid lines in this figure correspond to the results obtained for the $V_{lowk}$ interaction using the Wood Saxon wavefunctions of nuclear system representing $^{16}$O with variable size and are identical to the corresponding results in the left and middle panel of Fig.~\ref{fig4}.  Now, ignoring the contributions of $V_{lowk}$, which originates from partial waves with $L=1$ for the relative motion and total spin $S=1$  for the interacting nucleons, i.e. the partial waves $^3P_0$, $^3P_1$, and $^3P_2$, one obtains moderate modifications in the potential energy of the single-particle states, which are around 10 \% of the total contribution. These partial waves, however, are completely dominating the energy differences, which lead to the spin-orbit splittings. Therefore the results obtained for $V_{lowk}$ without the contributions of the $^3P_J$ partial waves, represented by the red dashed-dotted lines in Fig.~\ref{fig6} shows results very close to zero. 

This result may be used to conclude that the spin-orbit term in the non-relativistic shell-model of the nucleus originates from the spin-orbit structure in the two-nucleon interaction. This spin-orbit structure occurs in partial waves with orbital angular momentum $L\ge 1$ and spin $S=1$. The dominant contribution should occur in the $^3P_J$ partial waves while the effects in higher partial waves should be smaller due the finite range of the $NN$ interaction (see also \cite{zamick01}). In fact, this argument has been used e.g. to justify the origin of the spin-orbit term in simple phenomenological models for the effective $NN$ interaction like the Skyrme force\cite{VB72}, leading to the well known expression 
\begin{equation}
V_{ls}^{Skyrme} (r) = W \frac{3}{2}\frac{1}{r} \frac{d}{dr}\rho(r)\,,
\end{equation}
for the leading contribution to the spin-orbit term in the mean-field model for spherical nuclei with a density distribution $\rho(r)$ . This analytical form suggests that the strength of the spin-orbit splitting should scale with the size of the nuclear system $< r >$  like $< r > ^{-5}$. Indeed the dependence of the spin-orbit term evaluated from realistic $NN$ interactions, displayed  in Figs.~\ref{fig4} and \ref{fig5}, shows this scaling behaviour over a wide range of $< r >$. It is worth noting that the same scaling behaviour is also observed in Figs.~\ref{fig1} and \ref{fig3}, although the results presented in these figures are derived from the relativistic structure of the mean field in nuclear matter.

One of the most important results of relativistic models describing nuclear systems is the feature that a strong scalar component in the mean field of nuclei leads to an enhancement of the small component of the Dirac spinor representing the nucleon in the nuclear medium. This feature is represented by an effective Dirac mass $m^*$ for the nucleon, which is smaller than the nucleon mass $M$ in the vacuum. This implies that the matrix elements of a meson-exchange interaction of two nucleons in the nuclear medium should be evaluated for Dirac spinors with a reduced effective mass. In fact, Dirac Brueckner Hartree Fock calculation based on realistic OBE models for the $NN$ interaction have demonstrated that this effect is non-negligible and improves the results of calculations for bulk properties of nuclear matter and finite nuclei considerably\cite{Machxxx,brockm,samaru,ring17}.

Assuming a realistic value for the Dirac mass in the medium, like e.g. $m^*c^2$ = 600 MeV, one can calculate the matrix elements of the OBEPA for such Dirac spinors and evaluate a corresponding $V_{lowk}$ to represent the effective $NN$ interaction in the nuclear medium\cite{erik02}. Evaluating the spin-orbit splitting for this effective interaction one obtains the results represented by the blue dashed lines in Fig.~\ref{fig6}. Using the same Wood Saxon wavefunctions, i.e. the same radius $<r>$,  an enhancement of the spin-orbit splitting around 30 \% can be observed  (see also \cite{zamick02}).  

Rather similar results for this enhancement of the spin-orbit splitting due to the modification of the Dirac spinors as well as the dominance of the $^3P_J$ partial waves have also been observed for $^{40}$Ca (see Fig.\ref{fig5}).   

Since it has been demonstrated above that the results for the spin-orbit term are rather insensitive to the details of the nucleon wave functions, which yield the same radius for the nucleon distribution the remaining part of this section will consider oscillator functions, which are easily transformed from the partial waves for relative coordinates, which is used to evaluate the matrix elements for the $NN$ interaction, to oscillator functions in the coordinate system of the nucleus used to determine the properties of the nucleus. 

\begin{figure*}[htbp] \centerline{\includegraphics[width = 4.5in]{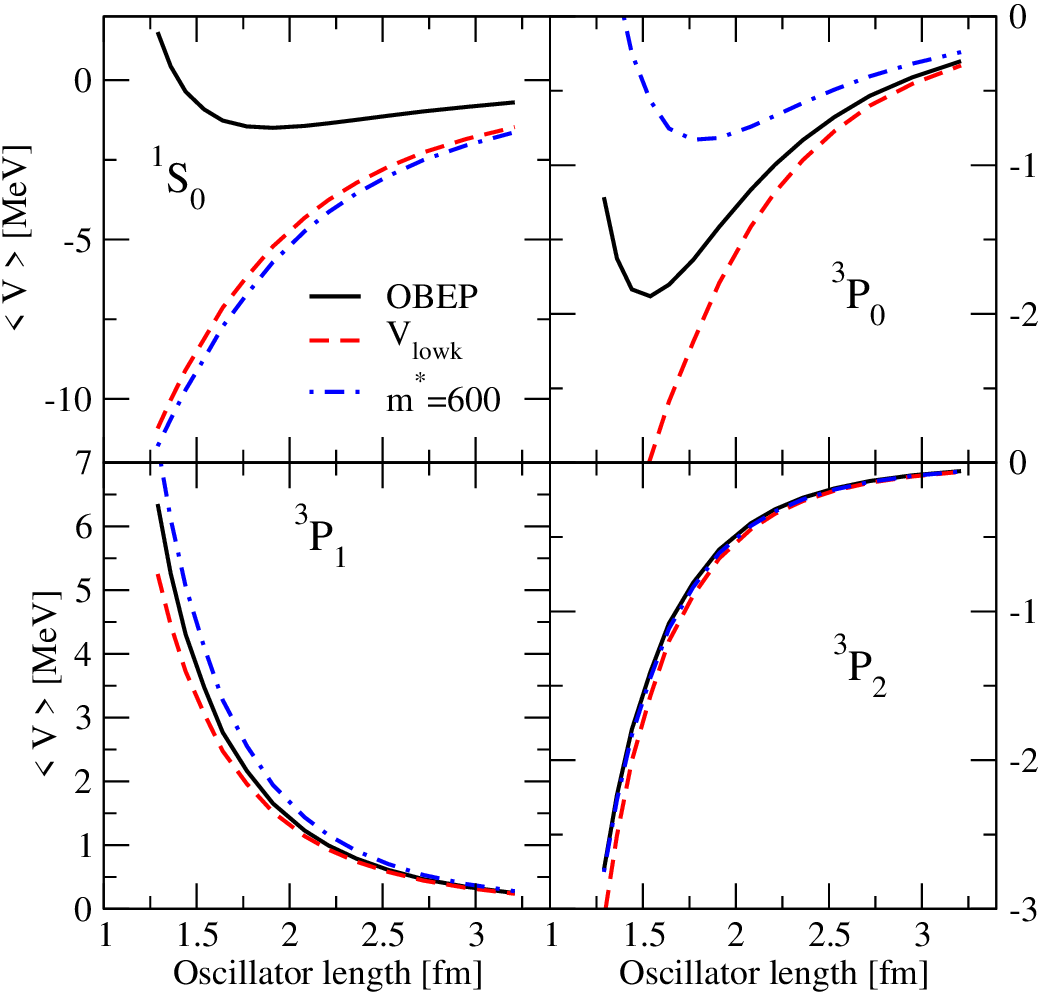}}
\caption{(color online) Results for diagonal matrix elements of the $NN$ interaction, $\langle n=0\vert V \vert n=0\rangle$, calculated in an oscillator basis as a function of the oscillator length. The four panels show results in different partial waves for the bare OBE potential and the resulting $V_{lowk}$ interaction. The dashed-dotted line (blue) represents result for the low-momentum interactions derived from the OBE potential for Dirac spinors with an effective Dirac mass $m^*$ of 600 MeV/c$^2$.}
\label{fig7}
\end{figure*}

Matrix-elements for the oscillator functions with radial quantum number $n=0$ are displayed in Fig.~\ref{fig7} as a function of the oscillator length 
\begin{equation}
b = \sqrt{\frac{\hbar}{M\omega}}\,,
\end{equation} 
with $M$ the mass of the nucleon and $\omega$ the oscillator frequency. The various panels show  the results for the partial waves $^3P_J$, which are relevant for the spin-orbit term in the nuclear shell model, as discussed above. For a comparison also matrix elements in the $^1S_0$ channel are given.  The matrix-elements of the bare OBE potential, represented by the black solid line are rather different from the corresponding results using $V_{lowk}$ (red-dashed lines) in the partial wave with orbital angular momentum $L=0$ the effects of the renormalization leading to $V_{lowk}$ are much weaker in the $^3P_J$ partial waves. This demonstrates that the renormalization accounts for high-momentum or short-range components of the underlying bare $NN$ interaction, which are strong in the $S$ channel and much weaker in partial waves with $L>0$.

The results displayed in Fig.~\ref{fig6} also show the effect of the reduced Dirac mass for the nucleons interacting in the nuclear medium by comparing matrix-elements of $V_{lowk}$ calculated for nucleon spinors of the vacuum to the low momentum interaction derived from the OBE interaction of nucleons with a Dirac mass $m^*c^2$ of 600 MeV.

One can see from this figure that the matrix-elements are attractive in $^3P_0$ and $^3P_2$ partial waves, whereas repulivse matrix-elements are obtained in the $^3P_1$ partial wave. Therefor it is obvious that the interaction in the various $^3P_J$ channels cannot simply be described in terms of a simple central plus a two-nucleon spin-orbit term. Other non-local components, like e.g. tensor- or quadratic spin-orbit terms, are very important part of a realistic $NN$ interaction. 

Assuming oscillator functions the matrix elements  displayed in Fig.~\ref{fig6} contribute to the spin-orbit splitting of the $0p$ shell in the nuclear mean field of $^{16}$O
\begin{equation}
\delta \varepsilon_p = 1.125 \langle V \rangle_{^3P_0} + 1.6875 \langle V \rangle_{^3P_1} - 2.8125 \langle V \rangle_{^3P_2}\,,\label{eq:partw}
\end{equation}
where $\langle V \rangle_{^3P_J}$ represents the relative oscillator matrix-element in the corresponding partial wave $^3P_J$ with radial quantum number $n=0$. Taking matrix-elements the corresponding panels of  Fig.~\ref{fig6} for an oscillator length, which is appropriate for $^{16}O$ ($b\approx 1.7$ fm), and applies eq.(\ref{eq:partw}) one finds that the negative contibution of the $^3P_0$ partial wave is more than compensated by the positive contributions originating from the $^3P_1$ and the  $^3P_2$ partial waves, leading to a total value around of 4 MeV, which is very close to the final result. 

The individual contributions originating from partial waves with $L=2$ are non-negligible. The corresponding matrix-element of the $^3D_1$ partial wave yields a contribution to $\delta \varepsilon_p$ of around 0.8 MeV. This energy shift, however, is compensated by the contributions from the other $^3D_J$ channels, which implies that the value of the spin-orbit splittings in $^{16}O$ is dominated by the contributions from $^3P_J$ partial waves (see also Fig.~\ref{fig6}).

The contributions of the $^3D_J$ partial waves is getting a bit more important in heavier nuclei with closed major shells, but also in $^{40}Ca$, the $^3P_J$ contributions are still dominant (see Fig.~\ref{fig5}). The situation is rather different in nuclei with closed subshells like $^{12}C$, $^{28}Si$ or $^{56}Ni$. The spin-orbit splittings in these nuclei obtain contributions also from other partial waves like $^1P_1$ or the tensor channel $^3S_1--^3D_1$, see discussion below. 

Results for selected spin-orbit splittings in various nuclei are listed in table~\ref{tab2} comparing various approximation schemes for the $NN$ interaction. Since the spin-orbit term is very sensitive to the radius of the nuclear system, the shell model wavefunctions have been fixed to oscillator functions, which yield a realistic value for the radius of the nucleus under consideration. The corresponding values for the oscillator frequency, $\hbar\omega$, are listed in the second line of this table.

The lines denoted as $V_{lowk}$. BHF, and OBEP, in the first column of table~\ref{tab2} present results of Hartree-Fock calculations assuming $V_{lowk}$ interaction, results of Brueckner Hartree Fock calculations, and Hartree Fock calculations assuming the bare OBEP A interaction defined in \cite{Machxxx}. Note that these different approaches yield results for the spin-orbit splitting, which are very close to each other, although individual single-particle energies and the total binding energies are rather different in these different approximation schemes. This confirms the finding above: The renormalization effects in $V_{lowk}$ and BHF to account for effects of short-range correlations have only little influence on the spin-orbit splitting, as the $ls$ term in the nuclear shell model arises from the $NN$ interaction in partial waves with $l=1$ and larger, which are not very sensitive to the treatment of short-range correlations. Therefore the subsequent discussion will consider modifications of the bare OBE potential.

It is remarkable that all 3 approaches yield large positive values for the spin-orbit splitting in the closed shell nuclei $^{16}$O and $^{40}$Ca, whereas small or negative values are obtained in nuclei with closed subshells. As discussed before, this indicates that the spherical shell model is not applicable in these cases. To explore the origin of this feature, the line in table~\ref{tab2} denoted as ''$T=0$'' shows the contribution of the $NN$ interaction for pairs of nucleons with isospin $T=0$. It is evident that the $T=0$ interaction leads to negligible contributions for the spin-orbit splitting of the closed shell systems, but provides negative contributions for the spin-orbit splitting of the nuclei with closed subshells. This supports the finding that the deformation of open shell nuclei mainly originates from the proton-neutron interaction.

Finally, the influence of the various mesons in the OBE potential shall be discussed. The line in table~\ref{tab2} marked as ''0.5*($\sigma,\omega$)´´ represents the results obtained with the OBEP quenching the contribution of the scalar $\sigma$ and the vector $\omega$ meson by a factor 1/2. Compared to the results using the full OBEP one finds a substantial reduction of the spin-orbit splittings. 

The matrix-elements of OBEP are most conveniently calculated in momentum space using ${\bf q}$ and ${\bf q'}$ to denote relative momenta for the pair of incoming and outgoing nucleons respectively. After transforming these expressions using momentum variables of an average relative momentum ${\bf p} = 1/2({\bf q} + {\bf q'})$ and momentum transfer ${\bf k} = ({\bf q'} - {\bf q})$ , one may expand the expressions in terms of ${\bf k}^2$ and ${\bf q}^2$. This leads to expressions for the $NN$ interaction with a two-body spin-orbit term, which has the same sign for the $\sigma$ exchange and the $\omega$ exchange (see \cite{Machxxx} and \cite{Mac86} for details). It is remarkable that the  $\sigma$ and $\omega$ mesons, which lead to contributions of opposite sign for the total energy, provide coherent contributions to the spin-orbit term. This is rather similar to the result obtained in the relativistic mean field (see eq.(\ref{eq15})).

Table~\ref{tab2} also shows results using the OBE potential, in which the contribution of the $\pi$-exchange is quenched by a factor 1/2 (see line with label $0.5*\pi$). One can see that the $\pi$-exchange has a negligible influence on the spin-orbit term in the case of the closed shell nuclei. It yields an attractive contribution in the case of open shell nuclei, in which the shell with $j=l+1/2$ is occupied whereas the one with $j=l-1/2$ is unoccupied. Note that the relativistic mean-field calculations discussed in first part of this section (see table \ref{tab1}) do not include pion exchange in the $NN$ interaction and yield positive spin-orbit splittings also for open shell nuclei.

This suppression of the spin-orbit term in open shell nuclei by the $T=0$ tensor interaction of a realistic $NN$ interaction has already been observed by Stancu et al.\cite{stancu77} in the attempt to include tensor components in the effective Skyrme interaction model. As the inclusion of these tensor components did not improve the results of mean field calculations they have been ignored for a long time. More recently, $T=0$ and $T=1$ tensor components in the Skyrme interaction have been adjusted to describe spin-orbit effects  in chains of isotopes of spin unsaturated nuclei (see e.g.\cite{stancu2,shihang19}). Note, however, that the parameters for these effective tensor terms are not deduced from a realistic $NN$ interaction.  
 
\begin{table}
\caption{\label{tab2} Spin-orbit splitting for various nuclei with closed shells or subshells calculated within an oscillator model with appropriate oscillator energies $\hbar\omega$ as listed in the first row. For the results displayed in the upper half of this table the OBE potential $A$ of \cite{Machxxx} has been used, while the results displayed in the lower part are based on the N$^3$LO interaction defined in \cite{Machn3lo}. All entries are given in MeV.} 
\begin{ruledtabular}
\begin{tabular}{c|ccccccc}
 &$^{12}$C &\multicolumn{2}{c}{$^{16}$O} & $^{28}$Si & \multicolumn{2}{c}{$^{40}$Ca }& $^{56}$Ni \\ 
$\hbar\omega$  &16.41 & \multicolumn{2}{c}{14.02} &11.95&\multicolumn{2}{c}{10.15}& 9.27\\
 & 0p & 0p & 0d &  0d & 0f & 1p & 0f \\
\hline
&&&&&&&\\
$V_{lowk}$ & -2.92 & 4.28 & 6.20 & -0.41 & 4.31 & 5.26 & 0.52 \\
BHF & -2.70 & 4.22 & 6.13 & -0.18 & 4.24 & 5.19 & 0.70 \\
OBEPA & -1.82 & 4.61 & 6.79 & 0.44 & 5.64 & 2.54 & 1.14 \\
T=0 &-2.79 & -0.02 & -0.02 & -1.71 & 0.06 &   0.03 & -1.20 \\
0.5*($\sigma,\omega$)& -3.89 & 2.69 & 4.23 & -1.31 & 3.68 & 1.78 & -0.38 \\
0.5*$\pi$ &2.29 & 4.61 & 6.51 &1.27 & 5.18 & 2.18 &2.75 \\
&&&&&&&\\
\hline
&&&&&&&\\
N$^3$LO, V & -3.01 & 4.26 & 6.16 & -0.64 & 5.19 & 2.29 & 0.30 \\
N$^3$LO + 3N, V & -2.37 & 5.72 & 7.98 & 0.21 & 6.77 & 2.90 & 1.17 \\
N$^3$LO, G & -2.81 & 4.07 & 5.93 & -0.41 & 4.99 & 2.20 & 0.46 \\
N$^3$LO + 3N, G & -2.08 & 5.34 & 7.51 & 0.45 & 6.34 & 2.72 & 1.31 \\
&&&&&&&\\
\end{tabular}
\end{ruledtabular}
\end{table}

All calculations discussed so far are based on the meson-exchange model for the $NN$ interaction. For a comparison a realistic $NN$ interaction shall be considered, which has been developed within the framework of chiral effective field theory by Entem et al.\cite{Machn3lo} They included term up to fifth order (N$^4$LO) and adjusted the parameter to provide very accurate fits for the $NN$ phase shifts and the data of the deuteron. Here, the model including terms up to fourth order (N$^3$LO) will be considered using a cutoff parameter $\Lambda$ of 450 MeV, which has also been defined in \cite{Machn3lo}.

\begin{figure*}[htbp] \centerline{\includegraphics[width = 4.5in]{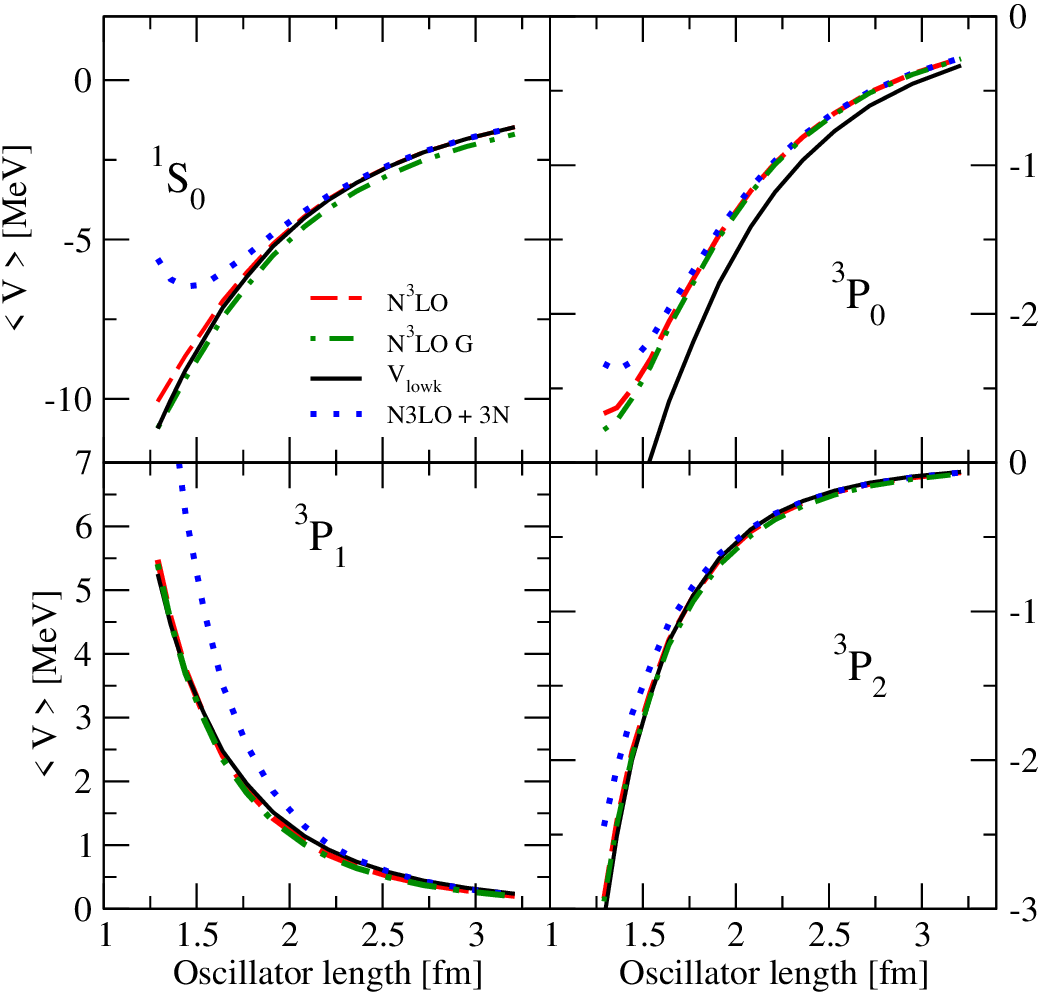}}
\caption{(color online) Results for diagonal matrix elements of the $NN$ interaction, $\langle n=0\vert V \vert n=0\rangle$, calculated in an oscillator basis as a function of the oscillator length. The four panels show results in different partial waves for the bare N$^3$LO interaction (red dashed line), the corresponding G matrix (green dashed-dotted line) and the N$^3$LO interaction including effects of a chiral 3N force expressed in form of a density-dependent effective $NN$ interaction (blue dotted line).  For a comparison also the results of $V_{lowk}$ (black solid line) have been copied from Fig.7.}
\label{fig8}
\end{figure*}

In anlogy to to Fig.~\ref{fig7} the Fig.~\ref{fig8} displays oscilator matrix elements in various partial waves. The difference between the the bare N$^3$LO potential (red dashed line) and the corresponding matrix elements of $G$ (green dashed dotted line) is very small even in the $^1S_0$ partial wave, indicating that the interaction is much softer than the OBE potential. The agreement between the matrix elements of the bare N$^3$LO interaction and the corresponding G matrix elements is of course even better in the $^3P_J$ partial waves, which makes it difficult to distinguish between the various lines. This is also reflected by the results for the spin-orbit splitting presented in the lower part of table \ref{tab2}. Results obtained in the Hartree-Fock approach (denoted by $N^3LO$, V) and the Brueckner Hartree Fock approacch ($N^3LO$, G) are almost identical.

For a comparison the Fig.~\ref{fig8} also includes matrix elements of the $V_{lowk}$ interaction already displayed in Fig.~\ref{fig7}. The agreement between the matrix elements derived from N$^3$LO and $V_{lowk}$ is good with small deviations in the $^3P_0$ partial wave. The same is true also for the calculated spin-orbit splittings presented in table 2. This demonstrates that nuclear structure calculations using $NN$ interactions, which give a very accurate fit of the $NN$ scattering phase shifts, lead to very similar results for the spin-orbit splitting, if the calculated nuclear radius is the same.

The chiral perturbation expansion also provides a consistent approach for $NN$ and many-nucleon interactions. In fact, 3N interactions are essential to provide a realistic result for the saturation point caculated for infinite nuclear matter. Without such a 3N interaction the chiral $NN$ interactions predict a saturation point of nuclear matter at very high density and very large binding energy, which is typical for soft $NN$ interactions, or no saturation at all. The chiral 3N interaction is usually represented in terms of a density-dependent $NN$ interaction\cite{NorbertK1,NorbertK2}, taking into account that it originates from a 3N interaction, when the Hartree-Fock mean field is calculated\cite{hebschw2010,carpols2013}.  

This representation of the chiral 3N interaction in terms of a density-dependent 2N interaction has also been used in the present study to investigate the influence of the chiral 3N interaction on the evaluation of the spin-orbit splitting in finite nuclei. Results for the relevant oscillator matrix elements are also displayed in Fig.~\ref{fig8} (blue dotted lines). Note, that the density parameter in the representation of the 3N force has been adjusted to the mean-value of the nucleon density distribution in the oscillator model for $^{16}$O, using the oscillator length $b$ displayed on the horizontal axis in this figure. This density parameter decreases with increasing $b$ and provides a reason for the fact that the difference between the matrix elements evaluated for N$^3$LO without and with inclusion of the 3N term vanishes with increasing $b$.

The effect of the 3N interaction is repulsive in all partial waves displayed in Fig.~\ref{fig8}. It is large in the $^1S_0$ partial wave, which is irrelevant for the spin-orbit term, and the $^3P_1$ channel. Note that additional repulsion in the $^3P_1$ partial wave enhances the spin-orbit splitting in closed shell nuclei (see discussion of eq.(\ref{eq:partw}) above). 

Indeed, the results displayed in table \ref{tab2} show a significant enhancement of the spin-orbit splitting for the closed shell nuclei. In the case of $^{16}$O this enhancement improves the agreement with the experimental splitting of 6.3 MeV for the hole states $p_{1/2}$ and $p_{3/2}$. On the other hand, however, the enhancement of the spin-orbit spliting for the particle states ($d_{3/2}$ and $d_{5/2}$) due to the 3N makes the agreement with experiment (5.08 MeV) even worse, as already the results from just 2N interaction overestimates this datum.

Here one must notice, that the density parameter used in the evaluation of the effective 2N interaction has been fixed to a value, which corresponds to the mean value of the nuclear density for the nucleus under consideration. This may be appropriate for the spin-orbit splittings of the hole states, but will overestimate the effect for the particle states. For the spin-orbit splitting of the particle states a smaller value for the density parameter would be appropriate or one may even take the limit $\rho =0$, which means to ignore the effect of the 3N interaction. Also one should keep in mind, that results displayed in table \ref{tab2} approximate the single-particle wave functions by oscillator functions. This is reasonable for the hole states but tends to overestimate the splitting for the particle states (see discussion above).

Note that this density-dependence of the $NN$ interaction resulting from a 3N interaction has been used by Nakada and Inakura\cite{nakada} to motivate a density-dependent spin-orbit term leading to a better description of isotope shifts in Pb nuclei.

In this context it is worth mentioning, that also the effects of effective Dirac mass $m^*$, discussed above, can be considered as a density dependence of the $NN$ interaction or as a 3N force described by the Z-diagram with intermediate hole states in the Dirac see. In the present study, which aims to explore the qualitative features leading to the spin-orbit splitting, the structure of the Dirac spinors in the nuclear medium have been represented by a global mass parameter. This may be appropriate for the spin-orbit splitting of the hole states and leads to agreement with experiment (see right panel of Fig.~\ref{fig6}) whereas for the particle states the assumption $m^*=m$ is preferable and leads to agreement with exeriment for the $d$-states (left panel of Fig.~\ref{fig6})\cite{zamick02}. 

\section{Summary}
The aim of this study has been to explore the occurrence of the spin-orbit term in the mean field of finite nuclei in the transition from infinite nuclear matter to finite nuclei. For that purpose sets of quasi-nuclear systems have been considered, describing nuclei with closed shells and variable size. Relativistic mean field calculations as well as non-relativistic approaches based on realistic models for the $NN$ interaction have been used to determine the spin-orbit splitting in the single-particle spectrum. 

One finds a very strong sensitivity of the results on the radius of the nuclear mass distribution. This means that results for the single-particle spectrum of nuclear systems should always be discussed together with the predicted radius.  The details of the underlying wavefunctions are not so important. Eigenfunctions of a Wood Saxon or oscillator potential as well as self-consistent Hartree-Fock wavefunctions yield almost identical results if they describe the nucleus with the same radius.

In the framework of relativistic, local mean field calculations this strong dependence on the size of the system can easily be understood. It is well known that a strong  spin-orbit term is obtained from the radial derivative of the difference between the scalar field $U_s$ and the vector field $U_0$.  Since $U_s$ is attractive, while $U_0$ is repulsive they cancel each other to a large extent in calculating binding energies. As the spin-orbit term results from the slope of the difference of $U_s$ and $U_0$ the potentials add up coherently.

In non-relativistic mean field calculations, which are based on $NN$ interactions fitting the $NN$ scattering data, the origin of the spin-orbit term in the nuclear mean field of closed shell nuclei can be found in the spin-dependence of the $NN$ interaction or scattering data. The studies show that the spin-orbit term mainly originates from the $NN$ interaction in partial waves with an orbital angular momentum $L=1$ for the relative motion and total spin $S=1$ of the interacting nucleons. The rules of anti-symmetrization require that the nucleons in these $^3P_J$ partial waves have a total isospin of $T=1$.

Effects of renormalization of the $NN$ interaction to account for the effects of short-range or high momentum component are not so important in the partial with $L\ge 1$ as in those with $L=0$. This means that calculated single-particle energies are rather different using a bare One-Boson-Exchange (OBE) potential or a renormalized interaction as $V_{lowk}$ or the Brueckner $G$-matrix, whereas the difference of the energies of spin-orbit doublets are almost identical calculated in terms of OBE, $V_{lowk}$ or using $G$ in the Brueckner Hartree Fock approximation.

The dominance of the partial waves with $L=1$ and spin of the interacting nucleons $S=1$ has been used to derive the spin-orbit term in a simple phenomenological model for the effective $NN$ interaction like the Skyrme force\cite{VB72}, leading to an expression, which suggests that the strength of  the spin-orbit splitting should scale with the size of the nuclear system $< r >$  like $< r > ^{-5}$. Indeed, the dependence of the spin-orbit term evaluated from realistic $NN$ interactions or derived from relativistic mean field calculations shows this scaling behaviour over a wide range of $< r >$. 

Using the meson-exchange model to describe the $NN$ interaction, one can identify the mesons which are responsible for the spin-orbit term. The spin-orbit term of closed shell nuclei mainly originates from the exchange of the scalar-isoscalar $\sigma$ meson and the vector-isoscalar $\omega$ meson. The contribution of these 2 mesons, which provide contributions to the binding energy with opposite sign, add up coherently in the spin-orbit term. This is in line with the observations of the relativistic mean field approach. It is also in agreement with the analysis of Machleidt\cite{Machxxx}, who pointed out that an expansion of the OBE amplitudes for $\sigma$ and $\omega$ exchange yields a spin-orbit term in the two-body interaction. This spin-orbit term in the $NN$ interaction could be thought to cause the $ls$ in the nuclear mean field. 
This line of argumentation is valid only in a rather qualitative way. Matrix-elements of realistic $NN$ interactions and phase shifts in the $^3P_J$ partial waves cannot simply be described in terms of a central and a two-body spin-orbit term.  

The analysis of the spin-orbit term is more complicate for nuclear systems with open shells. If a shell with $j=l+1/2$ is occupied whereas the corresponding one with $j=l-1/2$ is unoccupied in the spherical shell-model, the effects of the interaction in the $^3P_J$ channels get reduced and counterbalanced by effects of the interaction in the $T=0$ channel originating to a large extent from pion-exchange amplitudes. Within the spherical mean field approach this can even lead to negative values for the energy differences between the  states with $j=l-1/2$ and those with $j=l+1/2$. This means that the strong proton-neutron interaction favors deformed nuclei in this case.

The results on the spin-orbit term of nuclei, discussed so far are not limited to realistic meson-exchange potentials. It turns out that e.g. a realistic $NN$ interaction derived from chiral perturbation expansion (N$^3$LO defined in \cite{Machn3lo}) yields almost identical results and one may conclude that that nuclear structure calculations using $NN$ interactions, which give a very accurate fit of the $NN$ scattering phase shifts, lead to very similar results for the spin-orbit splitting, if the calculated nuclear radius is the same.

The chiral perturbation theory also provides a consistent description of 3 nucleon forces, which are frequently represented in terms of effective density-dependent $NN$ interactions.
The spin structure of this effective density-dependent interaction enhances the spin-orbit term considerably. Assuming that the relevant density is smaller for the evaluation of the spin-orbit term for the valence states above the Fermi surface than for the hole states, which are occupied, this can help to improve a simultaneous description of the spin-orbit term for particle and hole states.

The same is true also for the change of the Dirac spinors of the nucleons in the medium described in the framework of the Dirac Brueckner Hartree Fock approach in terms of an effective Dirac mass $m^*$. This effect can also be understood as an effective 3 nucleon interaction described in terms of a medium dependent 2 nucleon interaction which enhances the spin-orbit term.

The study of the spin-orbit term of the nuclear mean field presented in this work has been limited to the case of light nuclei with identical numbers for protons and neutrons ($Z=N$). It may be of interest to extend these studies to heavier nuclear systems to explore the isovector structure of the spin-orbit term more in detail.

\begin{acknowledgments}
In the early stage of this project I had various inspiring discussions with my late friend and colleague Arturo Polls. I want to thank him for this contribution as well as the many discussions, which we had over the years. Also I would like to thank Francesca Sammarruca and Ruprecht Machleidt for the access to the N$^3$LO interaction. 

 This work has been supported by the
the Deutsche Forschungsgemeinschaft (DFG) under
contract no. Mu 705/10-2.
\end{acknowledgments}

%\end{CJK*}
\end{document}